\documentclass[]{scrartcl}

\usepackage{graphicx}
\usepackage{caption}
\usepackage{subcaption}
\usepackage{siunitx}
\usepackage{xspace}
\usepackage{authblk}

\newcommand{\NV}{\ensuremath{\mathrm{NV}}\xspace}
\newcommand{\NVm}{\ensuremath{\mathrm{NV^{-}}}\xspace}

\newcommand{\CXII}{\ensuremath{\mathrm{^{12}C}}\xspace}

\title{Medical applications of diamond magnetometry: commercial viability}
\author{Matthew W. Dale and Gavin W. Morley}
\affil{Department of Physics, University of Warwick}

\begin{document}

\maketitle

\begin{abstract}

The sensing of magnetic fields has important applications in medicine, particularly to the sensing of signals in the heart and brain. The fields associated with biomagnetism are exceptionally weak, being many orders of magnitude smaller than the Earth's magnetic field. To measure them requires that we use the most sensitive detection techniques, however, to be commercially viable this must be done at an affordable cost. The current state of the art uses costly SQUID magnetometers, although they will likely be superseded by less costly, but otherwise limited, alkali vapour magnetometers. Here, we discuss the application of diamond magnetometers to medical applications. Diamond magnetometers are robust, solid state devices that work in a broad range of environments, with the potential for sensitivity comparable to the leading technologies.

\end{abstract}

\section{Introduction}

A magnetometer is a device that senses a magnetic field, measuring its strength and sometimes direction too. There are many different applications of magnetometry such as navigation and geo-surveying, but the one we focus on here is the measurement of medical biomagnetism, and specifically the magnetic fields produced by the heart.

\subsection{Magnetocardiography}

The cardiac cycle is initiated by an electrical impulse from the sinoatrial node, which then propagates through the heart. The electrical activity passes through specialised pathways, controlling the timing of the contraction of the tissue. The measurement of magnetic fields created by the heart is called magnetocardiography (MCG), and provides similar information to electrocardiography (ECG) which is widely used in hospitals. Several reviews of MCG are available, demonstrating that this technique provides diagnostic information additional to ECG for coronary artery disease and cardiac arrhythmias \cite{Kwong2013,Fenici2013,Fenici2005}. The MCG that has been carried out to date has used SQUID (superconducting quantum interference device) sensors at cryogenic temperatures. For this it is necessary to record magnetic fields with a sensitivity of \SI{50}{fT.Hz^{-1/2}} or better with a temporal resolution of around \SI{10}{\ms}. The frequency range from DC to \SI{50}{\Hz} is the most useful diagnostically, and frequencies from \SI{50}{\Hz} to \SI{250}{\Hz} are also useful. Existing SQUID-based MCG achieves sensitivity of better than \SI{10}{fT.Hz^{-1/2}} in a shielded room, or \SI{40}{fT.Hz^{-1/2}} in an unshielded room \cite{Fenici2013}. The installation of magnetically shielded rooms is expensive (up to \$1M with multiple shielding layers) and inconvenient for hospitals and the need for this would hamper attempts to bring MCG into common use. Existing SQUID-based shielded MCG systems cost around \$1M, but the next generation using diamond or alkali vapour cells should sell for significantly less due to the partial or total elimination of shielding and the costs related to cryogenics. Currently the non-industrialised cost for making SQUID sensors is about \$1k. The cost of new magnetic sensors will have to be competitive with SQUID sensors following industrialisation of the manufacturing processes. Detecting the smaller magnetic fields from human brains is also of interest, and this is called magnetoencephalography (MEG).

\subsection{Underlying physics of the negative nitrogen-vacancy centre in diamond}

Diamond is a remarkable material being best in class for many properties; it has excellent thermal conductivity, hardness and wide transparency to electromagnetic radiation to name a few. In recent years another application of diamond has emerged and that is as a sensor, owing to some unique properties of impurities that can be incorporated into the carbon lattice.

\begin{figure}
	
	\centering
	\begin{subfigure}[b]{0.3\textwidth}
		\includegraphics[trim=2cm 2cm 2cm 2cm, clip=true, width = \textwidth]{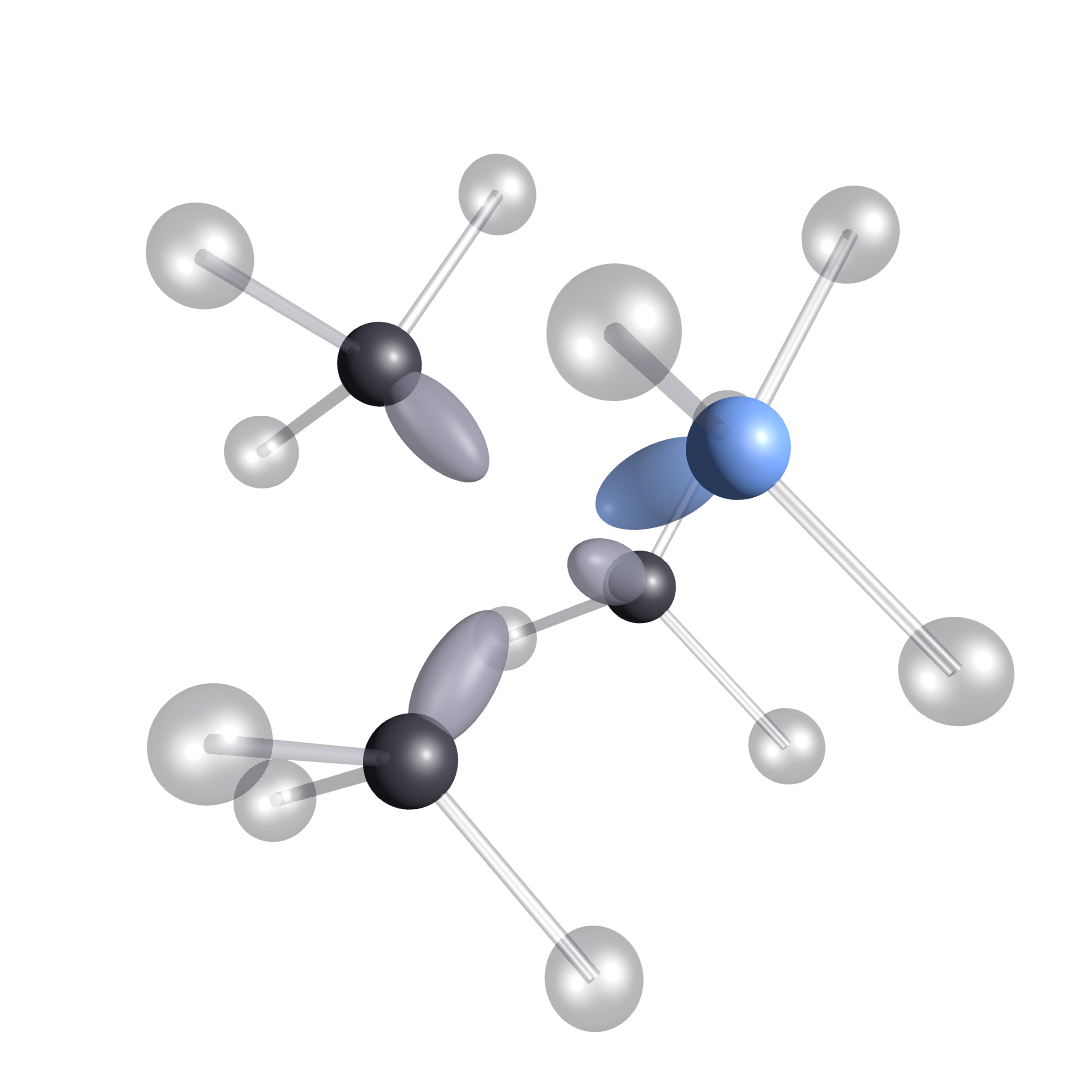}
		\label{fig:NV_sites:1}
	\end{subfigure}
	\qquad
	\begin{subfigure}[b]{0.3\textwidth}
		\includegraphics[trim=2cm 2cm 2cm 2cm, clip=true, width = \textwidth]{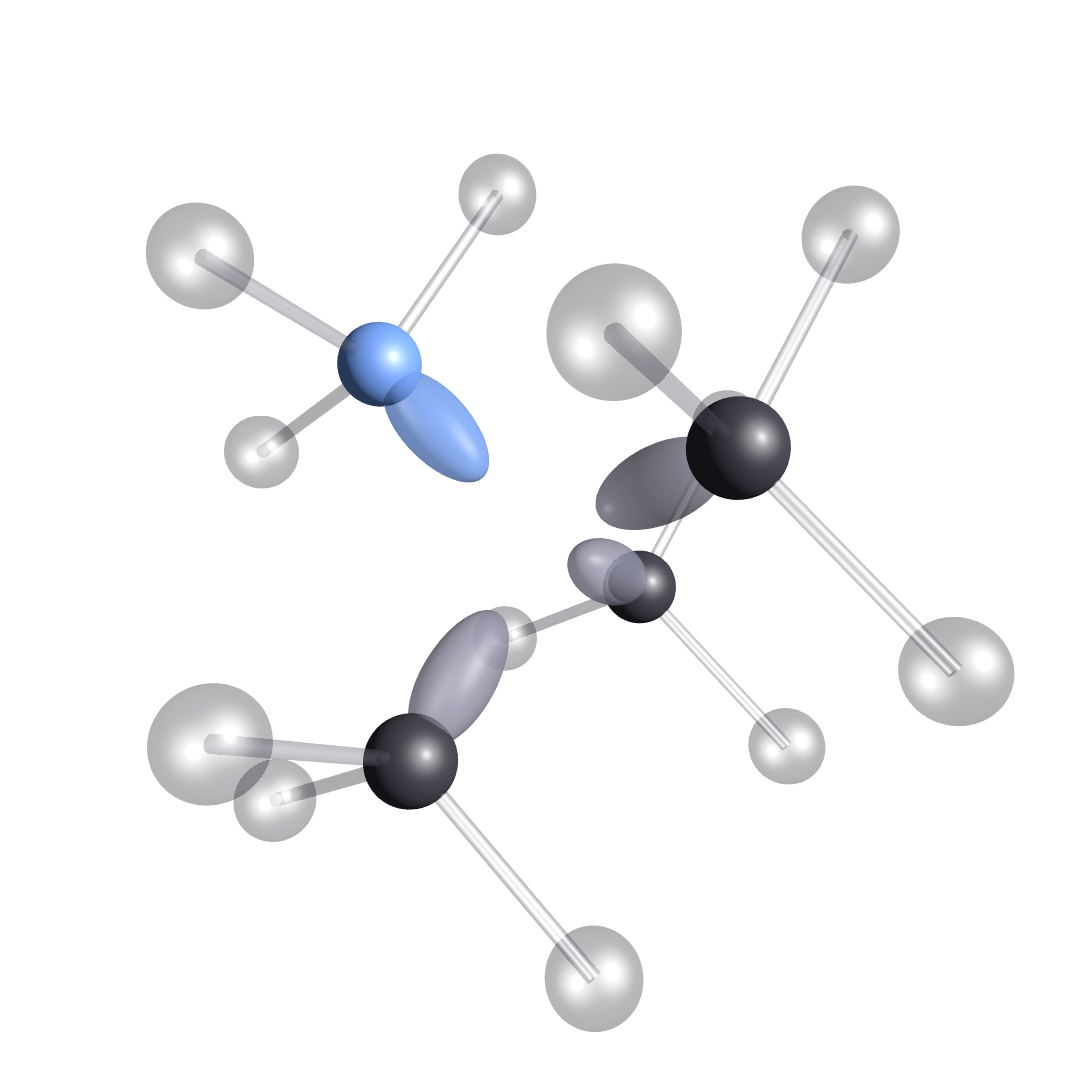}
		\label{fig:NV_sites:2}
	\end{subfigure}
	\qquad
	\begin{subfigure}[b]{0.3\textwidth}
		\includegraphics[trim=2cm 2cm 2cm 2cm, clip=true, width = \textwidth]{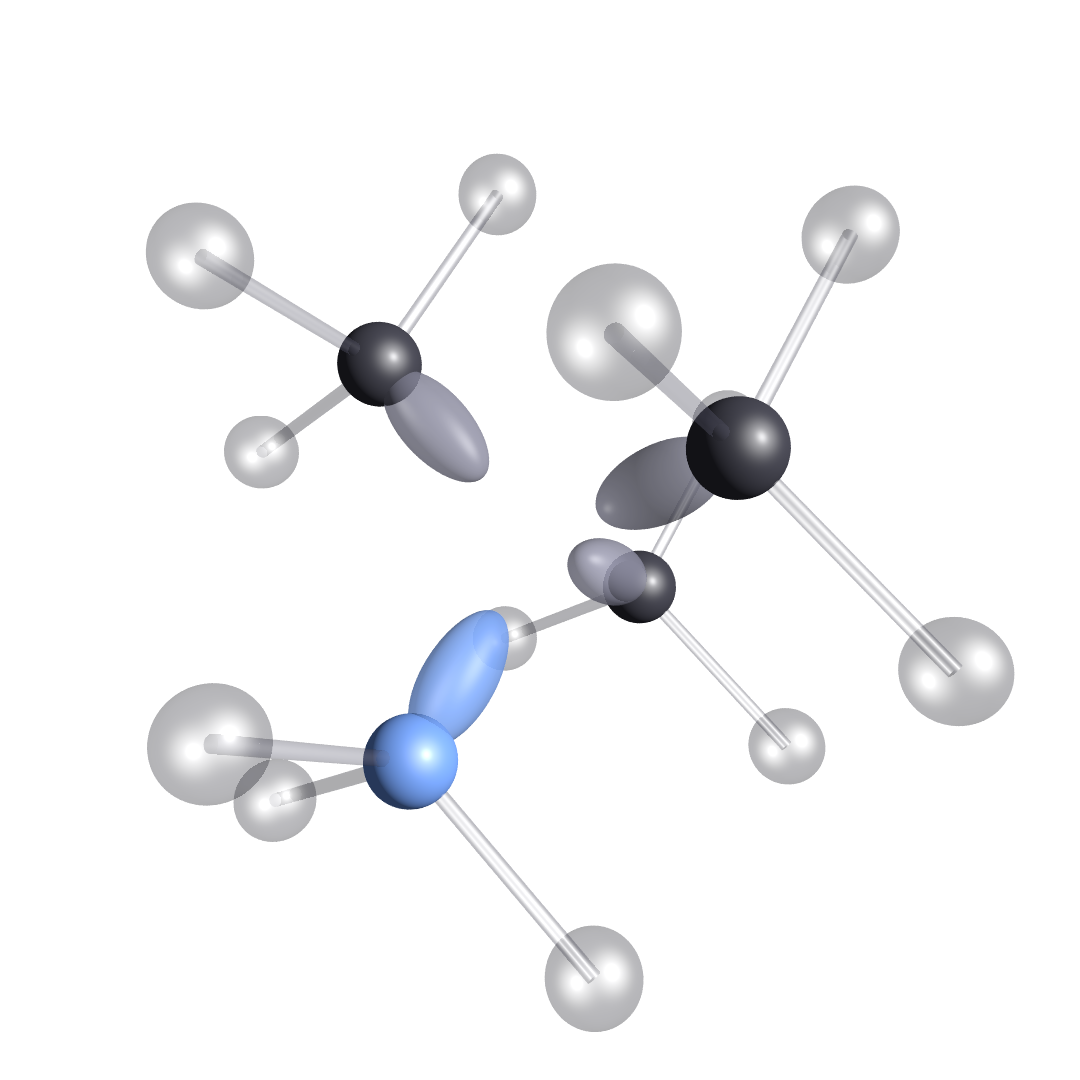}
		\label{fig:NV_sites:3}
	\end{subfigure}
	\qquad
	\begin{subfigure}[b]{0.3\textwidth}
		\includegraphics[trim=2cm 2cm 2cm 2cm, clip=true, width = \textwidth]{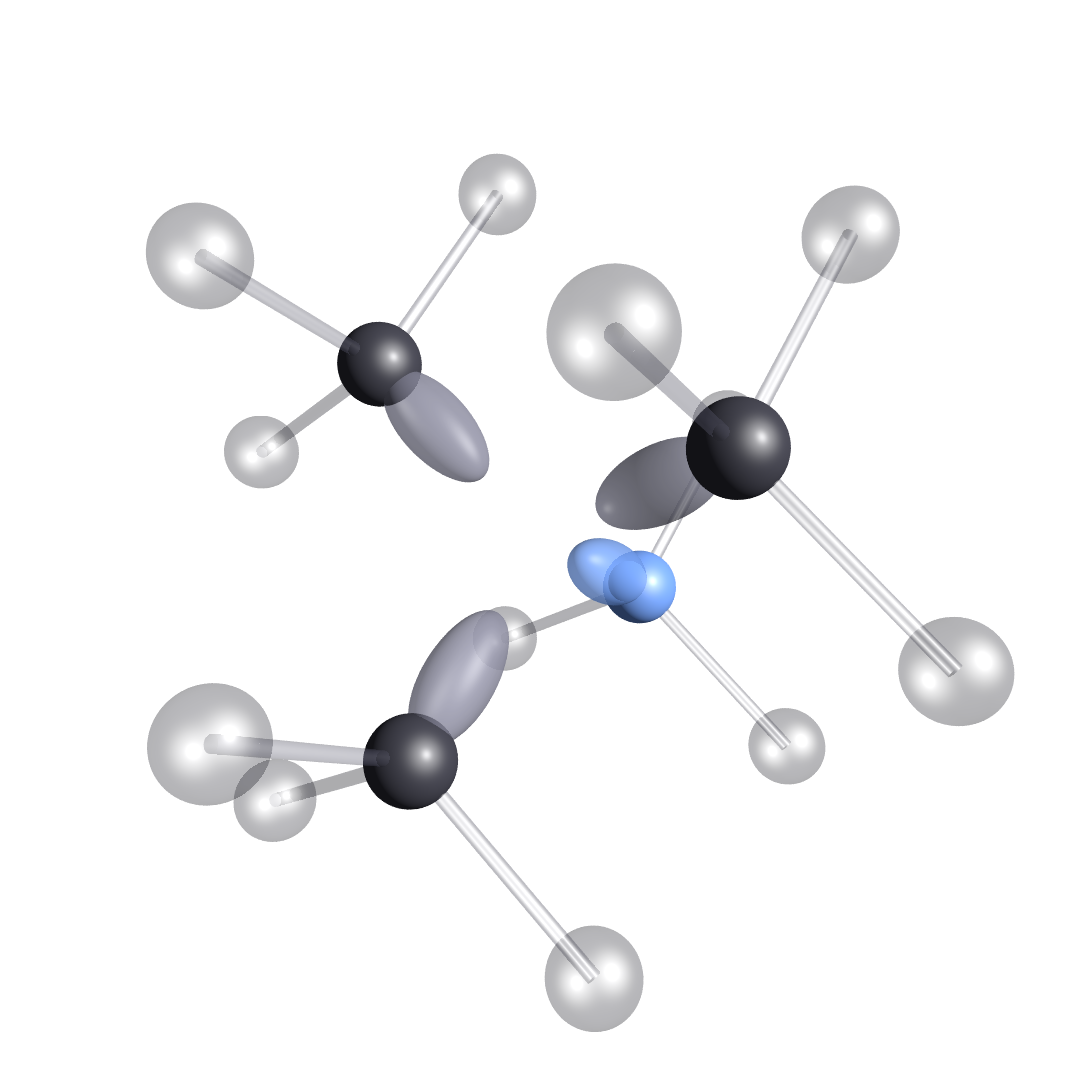}
		\label{fig:NV_sites:4}
	\end{subfigure}
	
	\caption{The nitrogen vacancy centre in diamond in its four orientations in the crystal lattice. The nitrogen atom is blue and the three carbon atoms surrounding the vacancy have been highlighted.}
	\label{fig:NV_sites}
	
\end{figure}

The band-gap of diamond is \SI{5.5}{\eV}, making it transparent to all visible light and into the UV, however impurities, or colour centres as they are called, introduce energy levels into the band-gap. Transitions involving these energy levels allow absorption and emission of visible light giving a diamond its colour. The most common colour centre in diamond is substitutional nitrogen, as it is abundant in the atmosphere and easily fits into the diamond lattice. Another is the vacancy, which may be incorporated during crystal growth as an imperfection, or introduced post growth by irradiation with high energy particles. If a vacancy becomes trapped adjacent to a substitutional nitrogen the nitrogen vacancy centre (\NV) is formed, a model of which is shown in Figure \ref{fig:NV_sites}. A substitutional nitrogen centre can donate an electron to NV to make it negatively charged (\NVm).

It is the \NVm centre which has some special properties and as such has been the subject of a considerable body of research. It is photostable and exhibits red photoluminescence that can be detected from a single centre \cite{Gruber1997}. The ground state has an unpaired electron spin that can be polarised optically with very high efficiency \cite{Doherty2011}. The same mechanism that is responsible for the polarization also reduces the luminescence intensity of the unpolarised states by up to \SI{30}{\percent}. These properties allow the optical detection of the spin of a single \NVm centre and consequently optically detected magnetic resonance (ODMR). An excellent review is given by Doherty $et$ $al.$ of the physics of the \NVm centre \cite{Doherty2013}.

A magnetic field splits the electron spin states, moving the position of the electron spin resonances, which can be read by ODMR. The accuracy of the field measurement is partly determined by the width of the resonance, which is primarily determined by the spin coherence lifetime. Diamond exhibits the longest coherence at room temperature of any solid state system \cite{Balasubramanian2009, Yamamoto2013a}.

The long coherence times stem from several properties of diamond. Firstly, the spin-lattice relaxation is small because of the weak spin-orbit interaction and strong covalent bonding of diamond. Secondly, decoherence due to interaction with other spins is minimal. At natural isotopic abundance diamond is \SI{98.9}{\percent} nuclear-spin free \CXII. In addition, there are few electron spins other than the \NVm centres we are interested in. The coherence times can be improved by engineering the diamond to remove parasitic defects \cite{Chu2014}, and enrichment with \CXII \cite{Yamamoto2013a}.

The \NVm centre has trigonal symmetry meaning that there are four possible orientations in which the N-V axis can sit in the diamond lattice; these are shown relative to each other in Figure \ref{fig:NV_sites}. The resonance frequency depends not only on the magnitude of the magnetic field, but also on the angle between the field and the N-V axis. Measuring the resonances from three or four sites of the \NVm centre allows vector measurements of the magnetic field.

\subsection{Nitrogen-vacancy sensitivity}

A review of \NVm magnetometry in the period up 2014 is available \cite{Rondin2014}. We should focus on demonstrated sensitivities rather than projected values. For bulk experiments, the best broadband sensitivity is \SI{15}{pT.Hz^{-1/2}}, from a diamond at room temperature with volume $13\times200\times2000$ \si{\micro \meter \cubed} for magnetic fields at frequencies from \SIrange{80}{2000}{\Hz} \cite{Barry2016a}. In a different experiment, using AC measurements instead of broadband (pulsed magnetic resonance instead of continuous-wave), the sensitivity to magnetic field at a particular frequency is higher, reaching \SI{0.9}{pT.Hz^{-1/2}} for a \SI{20}{\kilo\Hz} magnetic field with a bulk diamond \cite{Wolf2015}.

It has been proposed that using \NVm centres as a laser medium could provide significant further gains in sensitivity. By using the field dependence of the fluorescence to push the laser above threshold, contrast would be greatly increased. A so-called laser threshold magnetometer (LTM) based on \NVm is predicted to achieve a shot-noise limited DC sensitivity of \SI{2}{fT.Hz^{-1/2}} \cite{Jeske2016}.

\subsection{Current intellectual property holders}

There are a number of patents relating to diamond magnetometry, and in particular recently published by Lockheed Martin. Lockheed Martin patents cover vector magnetometry using ensembles of \NVm centres and a CW frequency swept experiment using a bias field to separate the spectrum \cite{Kaup2015}; a magnetometer using ensembles of \NVm centres with magnetic field detection by pulsed ODMR using an optimised Ramsey pulse sequence \cite{Egan2016}; a method for determining the orientations of NV centres to calibrate the NV magnetometer \cite{Manickam2016}; using parabolic or ellipsoidal reflectors to improve collection efficiency of luminescence and increasing sensitivity by resolving nitrogen hyperfine resonances \cite{Boesch2016}; and mico sized NV magnetometers \cite{StetsonJr.2016}.

Sensitivity to magnetic field can be increased by making improvements to the diamond material. Element Six have a number of relevant patents in this area, in particular a patent targeted at producing diamond for spintronic applications \cite{Scarsbrook2010}. The patent describes techniques for the growth of diamond by CVD with a low number of paramagnetic centres and \CXII isotopic enrichment.

A patent from Harvard describes improving the sensitivity of pulsed magnetometers by dynamical decoupling of electron spins from spin-spin interactions and interactions with the lattice \cite{Lukin2013}.

\section{Existing technologies}

There is a large array of technologies for magnetic field detection. Presently the most sensitive and applicable to MCG or MEG are the well-established superconducting quantum interference devices (SQUIDs) and atomic vapour cells made with alkali metals. Each of these technologies will be discussed in more detail here along with their strengths and weaknesses.

\subsection{SQUIDs}

Since the first reports of the Josephson effect \cite{Anderson1963} and SQUID magnetometry over 50 years ago \cite{Jaklevic1964} they have been the subject of extensive research, driven in part by their potential application to MEG \cite{Korber2016}. SQUIDs are one of the most sensitive magnetometers with typical sensitivities near \SI{1}{fT.Hz^{-1/2}} and a sensitivity of \SI{0.5}{fT.Hz^{-1/2}} at \SI{1}{\kilo\Hz} reported \cite{Fedele2015}. Current commercial MEG systems use arrays of SQUID magnetometers.

SQUIDs consist of a superconducting ring with one (RF SQUID) or more (DC SQUID) Josephson junctions. A small current less than a critical current can flow across the junction without creating a voltage. Because of flux quantisation, as the flux passing through the superconducting ring increases or decreases, the current flowing around it oscillates. When a bias current is applied this creates an oscillating voltage which can be read with conventional semiconductor electronics. The number of oscillations corresponds to the change in magnetic field.

The greatest limitation of SQUIDs is the requirement of cryogenic temperatures for operation. This increases cost and complexity and also reduces the sensitivity of a device since it must be some distance from the source of the fields being measured. In MEG the typical minimum distance between the SQUID sensor and brain is ~\SI{30}{\mm} \cite{Korber2016}. The distance can be decreased to increase SNR at the expense of helium boil-off. Furthermore, in current designs the sensors and Dewar are fixed meaning it is not optimised to head size, reducing sensor proximity.

SQUIDs have been demonstrated using high $T_C$ superconductors, with a sensitivity of ~\SI{4}{fT.Hz^{-1/2}}, allowing the use of nitrogen as a cryogen and closer sensor proximity \cite{Faley2013}, however there are significant challenges in their fabrication.

\subsection{Alkali metal magnetometer}

In the last decade the sensitivity of SQUID magnetometers has been challenged by atomic magnetometers (AM) using alkali metal vapours, which have the potential to replace SQUIDs in biomedical applications \cite{Shah2013}. Such AMs detect the Faraday rotation \cite{Budker2000} or absorption \cite{Shah2007} of light through a spin polarised vapour of potassium, rubidium or caesium. Their operation does not require the cryogenic cooling with helium required by SQUIDs, significantly reducing the complexity and cost of a device.

The sensitivity of an alkali-metal magnetometer is limited by spin relaxation time, to which the dominant contribution is spin-exchange collisions. For high density vapours and very low magnetic fields, the atoms can exchange spin much faster than the magnetic precession frequency. This is known as the spin-exchange relaxation free (SERF) regime \cite{Allred2002}, and allows sub \si{ft.Hz^{-1/2}} sensitivities. Sensitivities such as \SI{160}{aT.Hz^{-1/2}} with a measurement volume of \SI{0.45}{\cm \cubed} have been reported \cite{Dang2010}.


\section{Market}

There are around 100 SQUID MEG systems installed worldwide, at a cost of over \$1M each. The MCG market should be much larger if the instrumentation was affordable and portable, because MCG has been shown to be superior to ECG and hence other non-invasive approaches for the diagnosis of coronary artery disease (CAD) \cite{Kwong2013,Fenici2013,Fenici2005}. CAD is the most common type of heart disease and is the leading cause of death in the United States in both men and women. Several companies have tried and failed to commercialize SQUID-based MCG, held back by the cost of a cryogen-based system. We estimate that 100,000 MCG systems could be sold if the functionality were the same as existing SQUID systems and the price was below \$150k. This is based on there being over 100,000 hospitals in China, India, the EU, Japan and the USA.

Diamond magnetometers are at technology readiness level (TRL) 7: the technology has been demonstrated and is moving towards being put on sale. However, this has not yet reached the sensitivity needed for MCG, so an MCG system based on diamond is at TRL 4-5 (technology development).

\section{Outlook}

Although established SQUID sensors have the necessary sensitivity for MCG and MEG, the requirement for cryogenic cooling with helium makes them ultimately too expensive for widespread commercial applications. In addition, the need for the sensors to be in a Dewar limits how close they can be positioned to the subject.

Alkali metal vapour magnetometers offer the significant advantage over SQUID magnetometers of removing the need for cryogenic cooling. They do require heating, but this is much less challenging and restrictive. The most sensitive alkali metal vapour cells, operating in the SERF regime, however require being in very low field environments necessitating the use of magnetically shielded rooms. The background magnetic noise must be significantly reduced for all highly sensitive magnetometers, and this can be achieved to a reasonable degree by subtracting the signal from additional sensors slightly removed from the subject, however a magnetically shielded environment is still a requirement for SERF magnetometers.

Whilst it is true that the sensitivity of diamond \NVm magnetometers is presently several orders of magnitude worse than the state of the art SQUID or alkali metal magnetometers, the aforementioned technologies have been in development for significantly longer. As such it is very likely that through optimising diamond material, sensor configuration, and detection methodology (for example a pulsed microwave experiment rather than frequency swept), orders of magnitude improvements can be made. Even if \NVm magnetometers do not ultimately exceed others in sensitivity they might offer significant advantages in ruggedness, cost, and proximity to the subject.

A large part of the cost for both alkali metal and \NVm magnetometers is the laser excitation. For alkali metal magnetometers, atomic transitions are excited, requiring finely tuned and stabilised lasers. The \NVm transition which is excited is significantly broadened by coupling to phonons in the lattice, effectively allowing broadband illumination. With LED technology constantly improving, single colour LEDs are now available with powers of hundreds of milliwatts. The use of LED illumination rather than laser has the potential to dramatically reduce the cost of a device.

A challenge with \NVm magnetometers is that the signal is contained in luminescence from the diamond. Luminescence radiates in all directions making its collection difficult (this is not a problem with alkali metal magnetometers since they detect laser light transmitted though the vapour cell). However the use of ellipsoidal reflectors is discussed in a Lockheed Martin patent \cite{Boesch2016}, where it is stated that with correct implementation, near \SI{100}{\percent} collection should be possible. A nice design would be the use of an ellipsoidal reflector coupled to a fibre, such that all light transmitted through the fibre is focussed on the diamond, and all luminescence from the diamond is coupled in to the fibre.

\section{Acknowledgements}

We have benefited from useful discussions with Peter H{\"o}fer (Bruker), Paul Mawson (Texas Instruments), Riccardo Fenici (BACPIC), Donatella Brisinda (BACPIC), Ian Fisher and Tim LeClair. 


\bibliography{references}
\bibliographystyle{unsrt}

\end{document}